\documentclass{article}
\usepackage{PRIMEarxiv}

\usepackage{hyperref}
\usepackage[dvipsnames]{xcolor}
\usepackage{amsmath}
\usepackage{svg}
\usepackage{subcaption}
\usepackage{multirow}  % for multi-row tables
\usepackage{natbib}
\usepackage{graphicx}
\usepackage{gensymb}  % for using the degree symbol

\newcommand{\Fig}[1]{Fig.~\ref{fig:#1}}
\newcommand{\Tab}[1]{Table~\ref{tab:#1}}
\newcommand{\Sec}[1]{Section~\ref{sec:#1}}

\newcommand{\Eqn}[1]{Eqn.~\ref{eqn:#1}}

\newlength{\wholefigwidth}
\newlength{\smallfigwidth}
\newlength{\halfsmallfigwidth}
\newlength{\figwidth}
\begin{document}

% Title page stuff
\title{Spline-based neural network interatomic potentials: blending classical and machine learning models}

\author{
    Joshua A. Vita \\
    Department of Materials Science and Engineering \\
    University of Illinois Urbana-Champaign \\
    Urbana, Illinois 61801, USA
    \And
    Dallas R. Trinkle\thanks{Correspondence to: \url{dtrinkle@illinois.edu}}\\
    Department of Materials Science and Engineering \\
    University of Illinois Urbana-Champaign \\
    Urbana, Illinois 61801, USA
}

\maketitle
\begin{abstract}

While machine learning (ML) interatomic potentials (IPs) are able to achieve accuracies nearing the level of noise inherent in the first-principles data to which they are trained, it remains to be shown if their increased complexities are strictly necessary for constructing high-quality IPs.
In this work, we introduce a new MLIP framework which blends the simplicity of spline-based MEAM (s-MEAM) potentials with the flexibility of a neural network (NN) architecture.
The proposed framework, which we call the spline-based neural network potential (s-NNP), is a simplified version of the traditional NNP that can be used to describe complex datasets in a computationally efficient manner.
We demonstrate how this framework can be used to probe the boundary between classical and ML IPs, highlighting the benefits of key architectural changes.
Furthermore, we show that using spline filters for encoding atomic environments results in a readily interpreted embedding layer which can be coupled with modifications to the NN to incorporate expected physical behaviors and improve overall interpretability.
Finally, we test the flexibility of the spline filters, observing that they can be shared across multiple chemical systems in order to provide a convenient reference point from which to begin performing cross-system analyses.

\end{abstract}


\section{Introduction and Background}

In the fields of computational materials science and chemistry, machine learning interatomic potentials (MLIPs) are rapidly becoming an essential tool for running high-fidelity atomic-scale simulations.
% While the first MLIP to be successfully applied to practical systems was proposed using local environment descriptors and feed-forward neural networks \cite{Behler2007}, recent advancements in model architectures have enabled the development of models with accuracies nearing that of density functional theory (DFT) methods.
Notably, major breakthroughs in the field have typically been marked by the development of new methods for encoding the atomic environments into machine-readable descriptors, or using architectures from other fields of machine learning to improve regression from descriptors into energies and atomic forces.
The seminal work using atom-centered symmetry functions (ACSFs) \cite{Behler2007} as the encoder, followed by ``smoooth overlap of atomic position" (SOAP) descriptors \cite{Bartok2010}, and, most recently, equivariant message-passing networks \cite{Batzner2022} being milestones of particular importance in the field.
In parallel with these improvements to the encoding functions of MLIPs has been the development of new regression tools, favored either for the simplicity gained through the use of linear combinations of basis functions (e.g., \cite{Thompson2015,Shapeev2016,Drautz2019}) or the accuracy gained by increasing the effective cutoff radius of the model using message-passing neural networks (e.g., \cite{Gilmer2017,Schutt2018,Batatia2022}).
While countless other models have been proposed using combinations or variations of different methods (an incomplete list: \cite{Mueller2020,Manzhos2020a,Christensen2020a,Musaelian2022,Gasteiger2021,Haghighatlari2021,Lubbers2018,Hu2021forcenet,Batatia2022a}), the key insights remain the same: interatomic potentials can be greatly improved by leveraging architectures and optimization strategies drawn from other machine learning and deep learning applications.

Despite the success of these MLIPs, there has been a persistent need in the community for the continued development of low-cost classical IPs, particularly for use with large-scale simulations \cite{Sosso2016,Ravelo2013,Diemand2013,Phillips2020,ZepedaRuiz2017}.
Some of the major benefits that classical IPs \cite{Jones1924,Daw1984,Buckingham1938,Tersoff1986,Brenner2002,Shan2010,vanDuin2001,Baskes1992} have over MLIPs are their low computational costs, strong foundations in known physics, and a history of scientific research analyzing their behaviors and theoretical limitations.
Although improvements are being made to the speeds of MLIPs, the stark differences between classical and ML IPs in terms of size, design, and overall complexity make it difficult to leverage the well-established tools and knowledge from classical models in order to further improve their ML counterparts.

In this work, we develop a new IP model whose hyper-parameters can be tuned to transition smoothly between low-cost, low-complexity classical models and full MLIPs.
By basing the proposed model off of a classical spline-based MEAM potential, then extending it using a basic ML architecture, we enable direct comparisons to well-established classical forms as well as modern ML models.
These results help to bridge the gap between the two classes of models and highlight methods for improving speeds, interpretabily, and transferability of MLIPs.

\subsection{s-MEAM}
The model developed in this paper builds heavily upon the spline-based MEAM potential \cite{Lenosky2000} (``s-MEAM"), which we describe here in order to provide sufficient background to understand the new model architecture proposed in this work. The s-MEAM potential is a spline-based version of the popular analytical MEAM potential \cite{Baskes1992} that was intended to provide additional flexibility to the model while maintaining the same overall functional form. In the s-MEAM formalism, the energy $E_i$ of a given atom $i$ is written as:

\begin{equation}  \label{eqn:smeam}
    \begin{aligned} 
        E &= \sum_i E_i =  \sum_i \bigg[ U_{c_i}(n_i) + \sum_{j < i} \phi_{c_j}(r_{ij}) \bigg] \\
        n_i & = \sum_{j\neq i} \rho_{c_j}(r_{ij}) \\
        & \phantom{{}=1} + \sum_{\substack{j<k,\\j, k \neq i}} f_{c_j}(r_{ij})f_{c_k}(r_{ik})g_{c_jc_k}\!\left[\cos{(\theta_{jik})}\right].
    \end{aligned}
\end{equation}

In \Eqn{smeam} the energy of the $i$-th atom, $E_i$, is composed of a pair term ($\phi$) and an embedding energy contribution ($U$) for a given ``electron density" $n_i$, where all five functions ($\phi$, $U$, $\rho$, $f$, and $g$) are represented using cubic Hermite splines. The pair term is calculated by summing over pair distances $r_{ij}=|\vec{r}_j-\vec{r}_i|$ between each atom $i$ and its neighbors $j$ (with $r_{ij}$ less than a chosen cutoff distance, $r_c$). The electron density $n_i$ is further decomposed into 2-body ($\rho$) and 3-body (products of $f$ and $g$) contributions. The 2-body term is similar to the summation over $\phi$, while the 3-body term is computed by summing over the product of three spline functions that take $r_{ij}$, $r_{ik}$, and $cos(\theta_{jik})$ as inputs, where $\theta_{jik}$ is the bond angle formed by atoms $i$, $j$, and $k$ with $i$ at the center. The subscripts on the functions indicate that separate splines are used for evaluation depending on the chemistries $c_i$, $c_j$, and $c_k$ of atoms $i$, $j$, and $k$ (e.g., $g_{AA}$ for A-A bonds, $g_{AB}$ for A-B bonds, etc.).

In order to facilitate comparisons between s-MEAM and the model that will be proposed in this work, we will first define two new functions
\begin{equation}  \label{eqn:g}
    \begin{aligned} 
    G^\alpha_{3,i} (&r_{ij}, r_{ik}, \cos{\theta_{jik}}) = \\
    &\sum_{\substack{j<k \\ j,k \neq i}}  f^\alpha_{c_j}(r_{ij})f^\alpha_{c_k}(r_{ik})g^\alpha_{c_{jk}}\left(\cos{(\theta_{jik})}\right) \\
    G^\beta_{2,i}(&r_{ij}) = \sum_{j\neq i} \rho_{c_j}^\beta(r_{ij}),
    \end{aligned}
\end{equation}
then re-write \Eqn{smeam} as
\begin{equation} \label{eqn:smeam_rewrite}
    \begin{aligned}
        E & = \sum_i \bigg[ U_{c_i}(n_i) + \frac12 G^0_{2,i} \bigg] \\
        n_i & = \sum_{\beta=1}^{N_2} G^\beta_{2,i} + \sum_{\alpha=1}^{N_3} G^\alpha_{3,i} \\
    \end{aligned},
\end{equation}
where $N_2 = 1$ and $N_3 = 1$. We will henceforth refer to $G^\alpha_{3,i}$ and $G^\beta_{2,i}$ as 3-body and 2-body ``spline filters" respectively, in acknowledgement of the fact that they can be thought of as filters that characterize the local environment around atom $i$ in order to produce a scalar atomic environment descriptor $n_i$. In \Eqn{smeam_rewrite} we have introduced summations over the superscripts $\alpha$ and $\beta$ which currently only take on a single value of 1, but will be used later to denote different filters.

\subsection{NNP}

Shown to be universal function approximators \cite{Hornik1989}, neural networks (NNs) are provably more flexible than classical IPs which use explicit analytical forms.
Because of this, a sufficiently large NN would be expected to be able to accurately reproduce an arbitrary potential energy surface, assuming that it properly accounted for long-range interactions, was provided with enough fitting data, and did not suffer from limitations due to trainability.
The original Behler-Parrinello NNP \cite{Behler2007} was one of the first successful applications of NNs towards practical systems, where the atomic energy of atom $i$ is written as
\begin{equation}
    \label{eqn:nnp}
    \begin{aligned}
        E &= \sum_i N_{c_i}(\vec{D}_i)\\
        \vec{D_i} & = \langle D^1_{3,i}, \ldots, D^{N_3}_{3,i}, D_{2,i}^1, \ldots, D_{2,i}^{N_2} \rangle. 
    \end{aligned}
\end{equation}
In \Eqn{nnp}, $N_{c_i}$ is a neural network, $c_i$ is the element type of atom $i$, and $\vec{D}_i$ is the atom-centered symmetry function (ACSF) descriptor \cite{Behler2007} of atom $i$.
The ACSF local environment descriptor is comprised of radial symmetry functions
\begin{equation}
    \label{eqn:acsf_rad}
    D_{2,i}^\beta(r_{ij}) = \sum_{j \neq i} e^{-\eta^\beta (r_{ij} - R_s^\beta)^2} v_c(r_{ij}),
\end{equation}
parameterized by $\eta^\beta$ for changing the width of the Gaussian distribution, and $R_s^\beta$ to shift the distribution.
A smooth cutoff function $v_c(r_{ij})$ is used with the form:
\begin{equation}
    \label{eqn:nnp_cutoff}
    v_c(r_{ij}) = \bigg\{
    \begin{array}{lr}
        0.5 \times \big[ \cos{\frac{\pi r_{ij}}{r_c}} + 1 \big], & \text{if } r_{ij} \leq r_c\\
        0, & \text{if } r_{ij} > r_c.
    \end{array}
\end{equation}
Angular contributions are accounted for using the angular symmetry functions
\begin{equation}
    \label{eqn:acsf_ang}
    \begin{aligned}
        D_{3,i}^\alpha&(r_{ij}, r_{ik}, r_{jk}, \theta_{jik}) = \\
        &2^{1-{\zeta^\alpha}} \sum_{j,k \neq i}(1 + \lambda^\alpha \cos \theta_{jik})^{\zeta^\alpha} \\
        & \times e^{-\eta^\alpha (r^2_{ij}+r^2_{ik}+r^2_{jk})} v_c(r_{ij})v_c(r_{ik})v_c(r_{jk}).
    \end{aligned}
\end{equation}
Multiple radial and angular symmetry functions are constructed by making $N_2$ choices for $\eta^\beta$ and $R_s^\beta$, and $N_3$ choices for $\zeta^\alpha$, $\lambda^\alpha$ $(= \pm 1)$, and $\eta^\alpha$.
The evaluations of all of these symmetry functions are then concatenated together into a single vector $\vec{D}_i$ which is passed through a feed-forward neural network.

Obvious parallels can be drawn between the NNP form in \Eqn{nnp} and the re-written form of s-MEAM shown in \Eqn{smeam_rewrite}.
% In both cases, the atomic energy can be expressed as the evaluation of an embedding function operating on a local environment descriptor.
What differentiates NNP from s-MEAM, however, are the details regarding the construction of the local descriptors, and the form of the embedding function.
Where s-MEAM uses trainable spline filters for both the descriptor and the embedding function, NNP uses ACSFs and an NN respectively.
Although an NN would have an increased fitting capacity over the $U_{c_i}$ splines used in \Eqn{smeam_rewrite}, there are many similarities between the ACSF descriptors and the spline filters described in \Eqn{g}.
For example, the 2-body components of an ACSF descriptor, which are constructed by evaluating $D_{2,i}^\beta$ with multiple radial shifts $R_s^\beta$ for each neighboring atom $j$, can be viewed as basis functions used for interpolating the desired range of atomic distances.
This is conceptually related to how the basis functions of cubic Hermite splines allow $G_{2,i}^\beta$ to interpolate over its domain as well.
A similar argument can be made relating the angular components of ACSFs to the 3-body filters $G_{3,i}^\alpha$, where \Eqn{g} and \Eqn{acsf_rad} both multiply functions of pair distances ($f^\alpha_{c_k}(r_{ij})$ and $f^\alpha_{c_k}(r_{ik})$) by a function of the triplet angle ($g^\alpha_{c_{jk}}(\cos{(\theta_{jik})})$).
The ANI model \cite{Smith2017}, which will be used in this work for comparison to the model which we developed, is nearly identical to the NNP form described above, with the modifications that only a single $\eta^\beta$ is used, and \Eqn{acsf_ang} is altered to introduce both radial ($R_s^\alpha$) and angular ($\theta_s^\alpha$) shifts:
\begin{equation}
    \label{eqn:ani_ang}
    \begin{aligned}
        D_{3,i}^\alpha&(r_{ij}, r_{ik}, \theta_{jik}) = \\
        &2^{1-{\zeta^\alpha}} \sum_{j,k \neq i}[1 + \cos (\theta_{jik} - \theta_s^\alpha)]^{\zeta^\alpha} \\
        & \times e^{-\eta^\alpha (\frac{r_{ij}+r_{ik}}{2}-R_s^\alpha)} v_c(r_{ij})v_c(r_{ik}).
    \end{aligned}
\end{equation}

\section{Methods}

\subsection{s-NNP}
\label{sec:methods:snnp}

\begin{figure*}[!ht]
    \centering
     \includegraphics[scale=0.9]{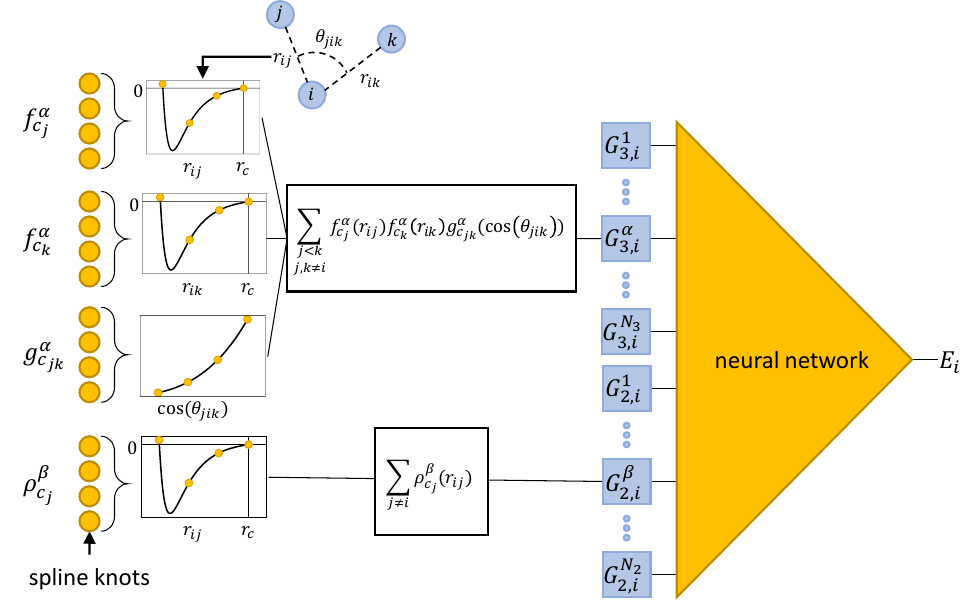}
     % \includesvg[width=\linewidth]{figures/main/snnp_model_architecture.svg}
    \caption{A diagram of the s-NNP model architecture. The spline knots parameterize individual 2-body ($f^\alpha_{c_j}$, $f^\alpha_{c_k}$, and $\rho_{c_j}^\beta$) or 3-body ($g^\alpha_{c_{jk}}$) splines, which take as input the pair distances $r_{ij}$, $r_{ik}$ or triplet angles $\theta_{jik}$ for all neighboring atoms $j$ and $k$ within a cutoff around each atom $i$. The outputs of the splines are then summed (after multiplying $f^\alpha_{c_j}$, $f^\alpha_{c_k}$, and $g^\alpha_{c_{jk}}$ together as described in \Eqn{g}) over all pairs or triplets. $N_3$ 3-body and $N_2$ 2-body filters are used (each parameterized by their own sets of knots, and indexed based upon the chemistries $c_j$, $c_k$, and $c_{jk}$ of each interaction), and the outputs are concatenated into a single vector which is propagated through a feed-forward neural network. All radial splines are pinned to have a value of zero at their cutoff distance. Unless otherwise specified, the s-NNP uses a CELU activation layer after every linear layer in the network except for the output layer.}
    \label{fig:model}
\end{figure*}

The main contribution of this work is the development of a spline-based neural network potential, outlined in \Fig{model}, which we call the ``s-NNP" (short for ``spline-NNP"). The s-NNP framework is intended to extend the fitting capacity of the s-MEAM class of potentials while maintaining the high interpretability and speed provided by the use of splines. s-NNP can be thought of as an s-MEAM potential with two critical changes: first, $N_3$ and $N_2$ in \Eqn{smeam_rewrite}, the numbers of $G^\alpha_{3,i}$ and $G^\beta_{2,i}$ spline filters, are taken to be hyper-parameters of the model; and second, the ``embedding" spline $U_i$ in an s-MEAM model is replaced by a fully-connected neural network. By including additional spline filters, the s-NNP is able to describe the local environment around an atom with increasingly fine resolution (since each spline can be thought of as a basis function for interpolating the atomic energies). The introduction of a neural network allows the model to achieve much more complex mappings into atomic energies than would be possible with the cubic splines $U_{c_i}$. In the s-NNP formalism, the energy $E_i$ of a given atom $i$ is then written as 

\begin{equation} \label{eqn:snnp}
    \begin{aligned}
    E & = \sum_i N(\vec{G_i}), \\
    \vec{G_i} & = \langle G^1_{3,i}, \ldots, G^{N_3}_{3,i}, G_{2,i}^1, \ldots, G_{2,i}^{N_2} \rangle. 
    \end{aligned}
\end{equation}

\noindent where $N$ is a neural network, and $\vec{G_i}$ is a vector of length $(N_3 + N_2)$. Notice that each component of $\vec{G_i}$ is computed by evaluating a different 3- or 2-body spline filter for the local environment of atom $i$. The parameters of an s-NNP that are trained during fitting are the positions of the knots of the $f^\alpha$, $g^\alpha$, and $\rho^{\,\beta}$ splines from \Eqn{g}, as well as the weights and biases in the neural network $N$. The hyper-parameters of the model are $N_3$, $N_2$, the number of knots in each spline, the number of layers in $N$, and the number of hidden nodes in each of those layers.

One benefit of the s-NNP framework is that it is closely related to both classical and machine learning interatomic potentials (see \Sec{other_models} for more discussion), making it possible to easily probe the performance gap between the two. For example, many classical potentials (LJ \cite{Jones1924}, EAM \cite{Daw1984}, MEAM \cite{Baskes1992}, Stillinger-Weber \cite{Stillinger1985}, Tersoff \cite{Tersoff1986}, and Buckingham \cite{Buckingham1938}) could be reformulated as s-NNP potentials with very few filters ($N_3 \in [0, 1]$, $N_2 \in [1, 2]$) and custom embedding functions instead of a neural network (though given the universal approximation theorem, these embedding functions could be represented using an NN as well). Because of this, we can easily construct spline-based ``classical" models by adjusting $N_3$ and $N_2$, but not including a neural network, then compare them directly to MLIPs by subsequently attaching networks with varying depths and widths. See \Sec{results} in the Results for details of such a study.

\subsection{Interpretability improvements}
\label{sec:interp}

\begin{figure}[!ht]
    \centering
     \includegraphics[scale=0.5]{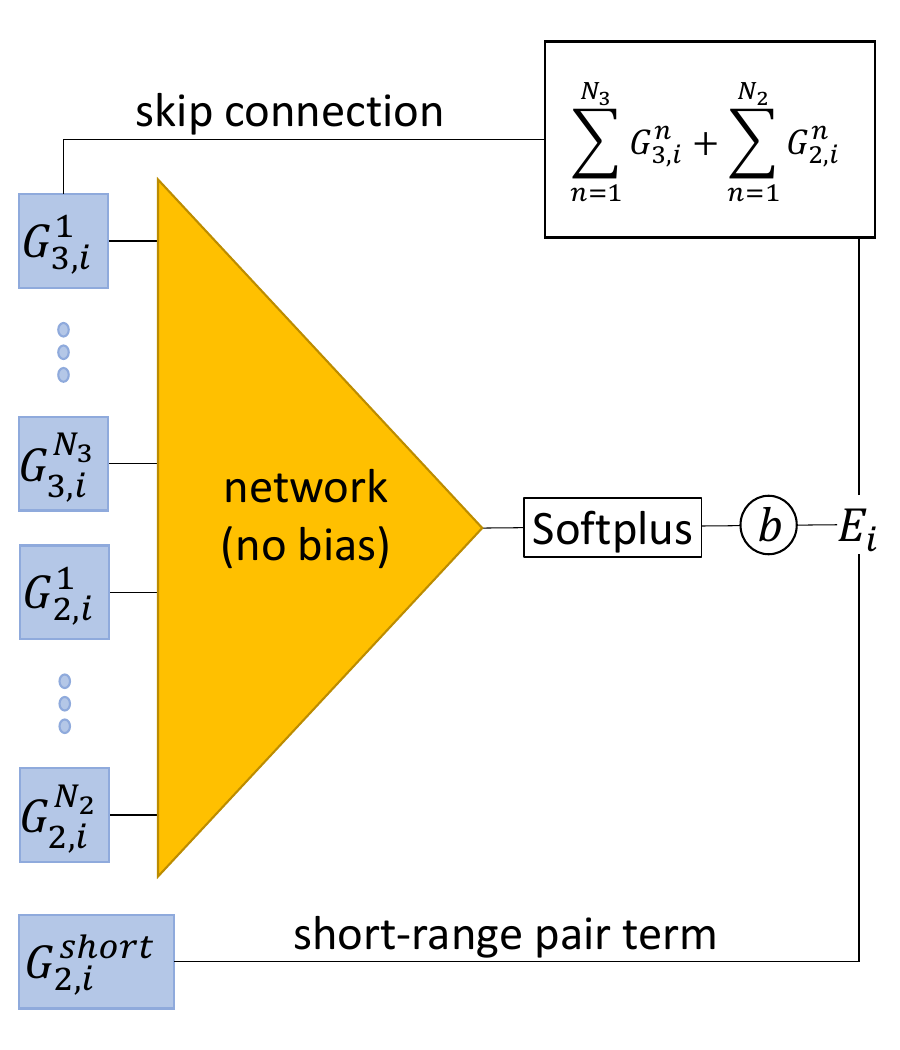}
    \caption{Modifications to the s-NNP form described in \Fig{model} for improving interpretability. The four modifications are: 1) adding skip connections; 2) removing the internal network bias, and adding an external bias, $b$; 3) wrapping the network outputs in a Softplus activation function; and 4) introducing a single short-range pair term using a 2-body filter. The short-range pair term is only allowed to be non-zero up to a cutoff of 2.5 \AA.}
    \label{fig:interp}
\end{figure}

A central tenet of constructing interpretable models is designing the model architecture in a way that helps isolate the contributions of specific parameter sets to the final model predictions.
With this in mind, we therefore propose the five modifications described in \Fig{interp}.
% , which will be discussed further in this section in order to highlight how they improve the interpretability of the model.
Although the modifications proposed here are only discussed in the context of s-NNP, they can also be applied to many other existing MLIP frameworks.

When applying all of the modifications described in \Fig{interp}, the full form of s-NNP is written as:
\begin{equation}
    \label{eqn:snnp_full_interp}
    \begin{aligned}
        E = \sum_i E_i &= \sum_i \bigg[ E_{\text{lin},i} + E_{\text{net},i} \bigg] \\
        E_{\text{lin},i} &= E_{\text{sc},i} + G^\text{short}_{2,i}+ b \\
        E_{\text{sc},i} &= \sum_{n=1}^{N_3} G^n_{3,i} + \sum_{n=1}^{N_2} G^n_{2,i} \\
        E_{\text{net},i} &=  \sigma\big[N_\text{no-bias}(\vec{G}_i) \big], \\
        % E_{\text{aux},i} &= G_{2,i},
    \end{aligned}
\end{equation}
where $E_{\text{sc},i}$ as a ``skip connection" term (as used in other DL fields \cite{He2015}), $G^\text{short}_{2,i}$ is a short-range 2-body filter, $b$ is a trainable isolated atom energy, $\sigma$ is the Softplus activation function $\sigma(x) = \log{(1 + e^x)}$, and $N_\text{no-bias}$ is a neural network with no bias terms on any of its layers.
Note that $E_{\text{lin},i}$ encompasses all of the terms which are linearly dependent upon the spline filters, and  $E_{\text{net},i}$ captures the non-linear dependence.
While a more in-depth discussion of the effects of each of these modifications can be found in \Sec{interp_discuss}, the practical result of \Eqn{snnp_full_interp} is that the spline filter visualizations like those shown in \Fig{filter_viz} can be intuitively understood.
For example, negative regions of the filters will usually correspond to negative $E_i$, and positive regions of the filters will always correspond to positive $E_i$.
While \Eqn{snnp_full_interp} still suffers from some drawbacks, predominantly associated with the non-linear effects of $N_\text{no-bias}(\vec{G}_i)$, we believe that it strikes a good balance between the accuracy achieved through a NN architecture, and the interpretability characteristic of classical potentials.

\subsection{Benchmarking dataset: Al}
\label{sec:datasets}
In order to test our framework, in \Sec{benchmarking} we fit s-NNP models to the aluminum dataset from Smith et al. \cite{Smith2021_al_al}, which serves as a good benchmarking dataset due to its size and configurational complexity. This dataset was built using an active learning technique for generating interatomic potential training data \cite{Smith2018_less_is_more}. It has also been shown to contain extremely diverse atomic environments, such that the ANI model \cite{Smith2019_ani} which was originally trained to the dataset was tested for use in shock simulations. After removing duplicate calculations (identical atomic positions and computed energies/forces, generated at different active learning steps), and one outlier configuration with particularly high forces, the final dataset consisted of 5,751 unique configurations (744,356 atoms total) with their corresponding DFT-computed energies and forces. A 90:10 train-test split was performed to partition the dataset into training/testing data. This data can be obtained from the original source \cite{ani_al_datasets}.

\subsection{Additional datasets: Cu, Ge, Mo}
\label{sec:flex_datasets}

As a test of the flexibility of the spline filters, in \Sec{flex} we trained an s-NNP model simultaneously to the Al dataset described in \Sec{datasets} and the Cu, Ge, and Mo datasets from Zuo et al. \cite{Zuo2020}. Each dataset from \cite{Zuo2020} was manually constructed and contains the ground state structure for the given element, strained supercells, slab structures, \textit{ab initio} molecular dynamics (AIMD) sampling of supercells at different temperatures (300 K and $0.5\times$, $0.9\times$, $1.5\times$, and $2.0\times$ the melting point), and AIMD sampling of supercells with single vacancies at 300 K and $2.0\times$ the melting point. On average, each training dataset includes approximately 250 structures for a total of 27,416 atoms (Cu), 14,072 atoms (Ge), and 10,087 atoms (Mo). A detailed summary of the contents of the datasets can be found in \cite{Zuo2020}. This data can be obtained from the original source \cite{mlearn_datasets}.

In order to attempt to balance the combined dataset, we took a random subset comprised of 20\% of the Al dataset in addition to the full Cu, Ge, and Mo datasets, resulting in a total training set size of 1271 Al configurations, 262 Cu, 228 Ge, and 194 Mo.
Another logical choice for constructing the multi-element training set would be to further downsample the Al dataset to better balance the relative concentrations of each element type.
However, we observed that doing so resulted in worse property predictions for Al, presumably due to the large number of high-energy configurations in the original Al dataset making random sampling of low-energy configurations relevant to the properties of interest unlikely.
In all cases, the atomic energies of each element type were shifted by the average energy of that type before training.

\section{Results} \label{sec:results}

\subsection{Benchmarking tests} \label{sec:benchmarking}

\begin{table*}[!ht]
    \centering
    \caption{
    Fitting results comparing linear s-NNP models, ``$(N_2,N_3)$ linear", s-NNP models with neural networks, ``$(N_2,N_3)$ $l=$*", s-NNP with interpretability improvements ``$(1,8)$ $l=5$, int. *", and the ANI model from \cite{Smith2021_al_al}, ``ANI". The $(N_2, N_3)$ notation for the spline layers indicates that $N_2$ 2-body and $N_3$ 3-body spline filters were used. Network architectures are denoted as a tuple of integers (``Network" column) specifying the number of nodes in each hidden layer of the model. Although the training/testing errors decrease significantly when adding additional splines to the linear models, their performance appears to begin to saturate with the ``(1,8) linear" model. Model performance only begins to be competitive with the ANI results upon the inclusion of a neural network with sufficient depth, which allows for non-linear combinations of spline outputs.
    While the ``(1,8) $l=5$, int. skip" model only adds the use of skip connections, the ``(1,8) $l=5$, int. all" and ``(1,8) $l=5$, int. wide" use all of the interpretability improvements discussed in \Sec{interp} and \Sec{interp_discuss}.
    Note that this means that the single 2-body filter specified using the ``(1,8)" notation is the short-range pair term described in \Fig{interp}, and that there are therefore no 2-body filters being passed through the network.
    Testing errors for the ``ANI" model were taken directly from \cite{Smith2021_al_al}, which did not report training errors. Note that the ANI model uses ensemble-averaging over 8 networks, which they report yields energy and force errors that are ``20\% and 40\% smaller, respectively, compared to a single ANI model".}
    \begin{center}
        \begin{tabular}{lcllllll}
            \hline\hline
            Model name & $(N_2, N_3)$ & Network & Parameters & \multicolumn{2}{c}{$E_{\text{RMSE}}$} & \multicolumn{2}{c}{$F_{\text{RMSE}}$} \\
            & & & (splines, network) & \multicolumn{2}{c}{(meV/atom)} & \multicolumn{2}{c}{(eV/\AA)}  \\
            & & & & Train & Test & Train & Test  \\ \hline
            $(1,1)$ linear & $(1,1)$ & - & (66, 0) & 74 & 78 & 0.76 & 0.82 \\
            $(1,2)$ linear & $(1, 2)$ & - & (110, 0) & 56 & 56 & 0.54 & 0.58 \\
            $(1,4)$ linear & $(1, 4)$ & - & (198, 0) & 40 & 41 & 0.46 & 0.49 \\
            $(2,4)$ linear & $(2,4)$ & - & (220, 0) & 38 & 36 & 0.46 & 0.50 \\
            $(1,8)$ linear & $(1, 8)$ & - & (374, 0) & 32 & 34 & 0.42 & 0.46 \\[5pt]
            
            $(1,2)$ $l=3$ & $(1, 2)$ & $(3,2,1)$ & (110, 23) & 44 & 44 & 0.33 & 0.35 \\
            $(1,4)$ $l=5$ & $(1, 4)$ & $(5,4,3,2,1)$ & (198, 80) & 10 & 11 & 0.22 & 0.23 \\
            $(2,4)$ $l=4$ & $(2,4)$ & $(6,3,2,1)$ & (220, 74) & 22 & 21 & 0.21 & 0.21 \\
            $(2,4)$ $l=5$ & $(2,4)$ & $(6,4,3,2,1)$ & (220, 96) & 8.7 & 9.0 & 0.19 & 0.20 \\
            $(2,4)$ $l=6$ & $(2,4)$ & $(6,5,4,3,2,1)$ & (220, 127) & 10 & 12 & 0.24 & 0.25 \\
            $(1,8)$ $l=5$ & $(1, 8)$ & $(9,8,4,2,1)$ & (374, 219) & 7.5 & 6.5 & 0.13 & 0.13 \\[5pt]
            % $(8,8)$ $l=5$ & $(8,8)$ & $(16,8,4,2,1)$ & (528, 457) & 5.7 & 4.7 & 0.19 & 0.19 \\
            % $(16,16)$ $l=5$ & $(16,16)$ & $[32,16,8,4,1]$ & (1056, 1761) & 4.5 & 3.5 & 0.14 & 0.15 \\
            % $(8,8)$ wide $l=5$ & $(8,8)$ & $(128,64,32,16,1)$ & (528, 13,057) & 5.0 & 4.2 & 0.17 & 0.18 \\[5pt]

            % $(2,4)$ $l=6$, skip & $(2,4)$ & $[6,5,4,3,2,1]$ & (220, 127) & 9.7 & 7.0 & 0.19 & 0.20 \\
            $(1,8)$ $l=5$, int. skip & $(1, 8)$ & $(9,8,4,2,1)$ & (374, 219) & 5.5 & 4.6 & 0.15 & 0.15 \\
            $(1,8)$ $l=5$, int. all & $(1, 8)$ & $(9,8,4,2,1)$ & (374, 219) & 5.5 & 5.9 & 0.12 & 0.12 \\
            % $(8,8)$ $l=5$, skip & $(8,8)$ & $(16,8,4,2,1)$ & (528, 457) & 5.6 & 4.0 & 0.13 & 0.14 \\
            % $(2,4)$ wide $l=5$, skip & $(2,4)$ & $[128,64,32,16,1]$ & (220, 11,997) & 8.7 & 7.6 & 0.17 & 0.18 \\
            % $(8,8)$ wide $l=5$, skip & $(8,8)$ & $(128,64,32,16,1)$ & (528, 13,057) & 3.8 & 3.0 & 0.10 & 0.11 \\
            % $(8,8)$ wide $l=5$, skip, $L_2$  & $(8,8)$ & $[128,64,32,16,1]$ & (528, 13,057) & 4.0 & 2.8 & 0.10 & 0.10 \\
            $(1,8)$ $l=5$, int. wide  & $(1,8)$ & $(128,64,32,16,1)$ & (374, 11,793) & 5.9 & 5.1 & 0.13 & 0.14 \\[5pt]
            
            ANI \cite{Smith2021_al_al} & - & $(96,96,64,1)$ & (-, 15,585) & - & 1.9 & - & 0.06 \\
             \hline\hline
        \end{tabular}
    \end{center}
    \label{tab:fits}
\end{table*}

Using the s-NNP framework, we fit a collection of models to the Al dataset to probe the effects of increasing model capacity in two distinct ways: first by increasing the number of spline filters, and second by increasing the size of the network used for mapping filter outputs to energies.
\Tab{fits} outlines the architectures and accuracies of the trained models, grouped by key architectural changes and sorted by total number of parameters.
The trained s-NNP models can be conceptually broken down into three main groups, each highlighting the effects of specific architectural changes: 1) ``$(N_2, N_3)$ linear" models, increasing the number of splines when no neural network is used; 2) ``$(N_2, N_3)$ $l=$*" models, increasing the number of splines and network size; and 3) introducing the interpretability improvements described in \Sec{interp}.

The results in \Tab{fits} show multiple avenues for systematically improving the performance of an s-NNP model, though each method appears to experience a saturation point beyond which the model suffers from diminishing returns as complexity increases.
Using the ``$(1,1)$ linear" model as a baseline, we see that increasing $N_3$ from 1 to 8 can monotonically decrease both the energy and the force errors. Increasing $N_2$ has no significant effect on the accuracy of the linear models, which is consistent with the notion that a linear combination of cubic Hermite splines can be represented using a single spline.

The introduction of a neural network enables the model to better utilize the additional spline filters, leading to significant improvement of the ``$(1,2)$ $l=3$" model over any of the linear models.
Increasing the number of spline filters, and subsequently increasing the network width and depth to maintain a ``funnel-like" structure (decreasing layer width with increasing depth) can also lead to steady improvements.
However, with increasing model size we began to see a commensurate increase in training difficulty, often resulting in larger models with higher errors than what might be expected based on the performance of their smaller counterparts (e.g., ``(2,4) $l=6$" compared to ``(2,4) $l=5$", or ``(1,8) $l=5$, int. wide" compared to ``(1,8) $l=5$, int. all").
Notably, the interpretability improvements from \Sec{interp} did not hurt model performance, demonstrating that accuracy and interpretability are not strictly opposing attributes of a model.
Based on the results shown in \Tab{fits}, we will use the ``(1,8) $l=5$, int. all" for all experiments and analyses in the remainder of this paper.
% Introducing skip connections allowed the training routine to find better minima, consistent with observations that skip connections improve trainability in other deep learning applications \cite{Li2017}.
% $L_2$ regularization (penalizing the sum of squared weights) was found to improve the transferability of the splines (see \Sec{flex}) without increasing errors on the Al dataset.
% We will therefore use the ``(8,8) wide $l=5$, skip, $L_2$" for all experiments and analyses in the remainder of the paper.
% Note that for the spline filters, the $L_2$ regularization is not applied directly to the spline knots, but instead to the coefficients of the piece-wise polynomials comprising the spline.

\begin{figure*}[!ht]
    \centering
    \includegraphics[width=\linewidth]{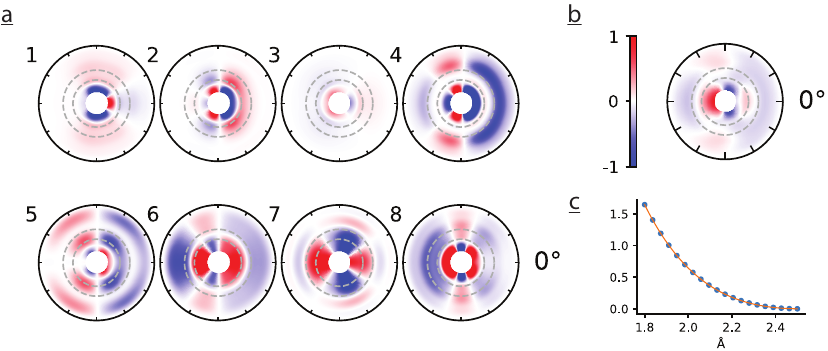}
    
    \caption{Visualizations of spline filters of the ``(1,8) final" model from \Tab{fits} for atomic distances in the range [2.5\AA, 7\AA]. a) Plots of the individual $G^\alpha_{3,i}$ filters, where integer labels above each plot indicate the index $\alpha$. b) The average of all $G^\alpha_{3,i}$ 3-body filters. c) The short-range pair term, as described in \Sec{interp}. Note that the radial splines $f_{c_j}^\alpha$, $f_{c_k}^\alpha$, and $\rho_{c_j}^\beta$ use linear extrapolation for distances outside of the domain defined by their knots. Dashed grey lines in the polar plots correspond to the first and second nearest-neighbor distances (2.86 \AA \space and 4.05 \AA \space respectively) for FCC Al at room temperature \cite{Nakashima2020}. Each point in the polar plots is computed by fixing atom $i$ at the origin, fixing atom $j$ at the given $(r, \theta)$, fixing atom $k$ along the $\theta = 0$ axis, then integrating $G^\alpha_3$ over $r_k \in [1.5, 7.0]$. Note that there is a forced symmetry in the polar plots since the $g^\alpha_{c_{jk}}$ splines in \Eqn{g} take $\cos{(\theta)}$ as input.
    All polar plots use the same color scale, where values are clipped to fall within a chosen range to optimize for visibility while avoiding information loss. Values for $r < 1.5$ \AA, which was the smallest distance sampled in the Al dataset as shown in \Fig{supp:rij_costheta_distributions}, are omitted to ensure that high signals at small atomic distances would not wash out the rest of the colors in the plots. Though all plots in this figure are technically in units of eV, this would not be true if skip connections were not used.}
    % Furthermore, while the integration over $r_k$ is useful for visualization, it results in artificially large values that would not be seen in practice.}
    \label{fig:filter_viz}
\end{figure*}

\subsection{Model visualization}
\label{sec:visualization}

A major advantage of s-NNP over many other MLIPs, especially when coupled with the modifications from \Sec{interp}, is that the spline filters $G_{3,i}^\alpha$ and $G_{2,i}^\beta$ lend themselves to easy visualization. This can be valuable for helping model developers and users to better understand how their model is interacting with the data or influencing simulation results.
The polar plots in \Fig{filter_viz}a, corresponding to the ``(1,8) $l=5$, int. all" model, represent the total activation of the 3-body filters induced by placing an atom at a given $(r, \theta)$.
These visualizations make it easy to recognize aspects of the local environments around an atom that are learned during training to have lower energy.
For example, the averaged filter shown in \Fig{filter_viz}b has an attractive behavior of the model for bond angles $60 \degree \leq \theta_{jik} \leq 90 \degree$ in addition to repulsion for angles larger than approximately $120 \degree$.
Many of the individual filters in \Fig{filter_viz}a also show characteristic features at various bond angles, especially for bond lengths near the first and second nearest neighbor distances in FCC aluminum.
\Fig{filter_viz}c shows the short-range pair term, which was learned to have a strongly repulsive contribution, as would be expected based on the Pauli exclusion principle.

It's worth mentioning that similar plots to the ones shown in \Fig{filter_viz} could also be generated for other NNP-like models, for example by visualizing each of the components of the vector output of the first hidden layer in an ANI model.
However, most other neural network-based architectures would have significantly more filters to visualize given the size of their hidden layers.
For example, the ANI model in \Tab{fits} has 96 nodes in its first hidden layer, thus making it more difficult to interpret the results.
Furthermore, since ANI (and most other models) does not use skip connections summing the hidden layer directly into the output, any resultant visualization will not necessarily be in units of energy, meaning it may undergo significant non-linear transformations as it passes through the network.

\subsection{Flexibility tests}
\label{sec:flex}

\begin{figure}[!ht]
    \centering
    \includegraphics[scale=0.625]{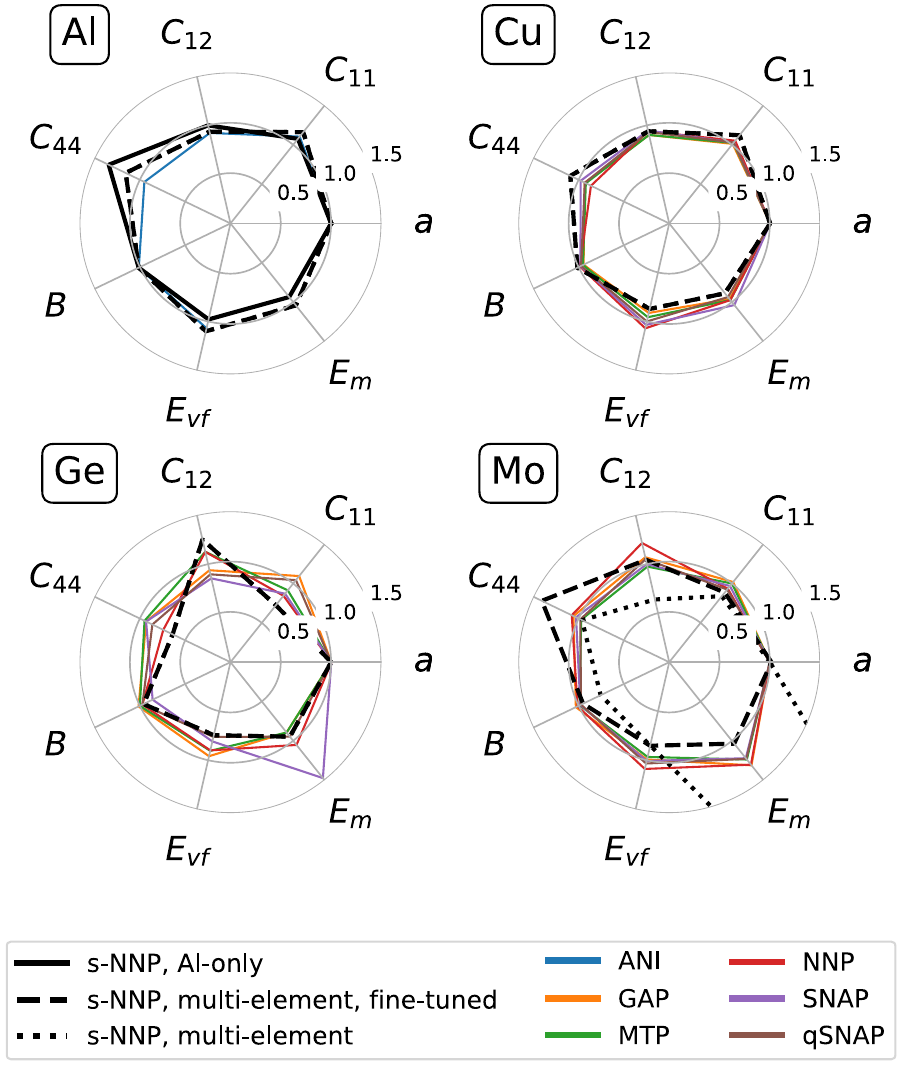}
    \caption{Property predictions of the lattice constant ($a$), cubic elastic constants ($C_{ij}$), bulk modulus ($B$), and vacancy formation and migration energies ($E_{vf}$ and $E_m$) for s-NNP models compared to existing MLIPs. s-NNP results are from this work, ANI results from \cite{Smith2021_al_al}, and all others from \cite{Zuo2020}. All properties have been normalized with respect to the DFT-predicted values. Note that \cite{Smith2021_al_al} did not compute $E_m$ for ANI. The ``(1,8) $l=5$, int. all" model (solid black line), which was trained only to Al, predicts all seven properties well, with the exception of $C_{44}$, which has a relatively small magnitude compared to the other elastic constants. The performance of the multi-element s-NNP model (dashed black lines) is within the range of the other single-element MLIPs, though s-NNP's predictions for Ge are somewhat distorted.
    The ``fine-tuned" s-NNP model was manually adjusted after training was completed in order improve the property predictions, as described in \Sec{flex}.
    }
    \label{fig:spider}
\end{figure}

\newcommand{\td}{ $|$ }

The high interpretablility of the s-NNP spline filters, as discussed in \Sec{visualization}, is particularly valuable when the filters can be applied to different chemical systems in order to enable cross-system comparisons.
For example, if multiple s-NNP models using the same spline filters were trained to different chemical systems, then the networks of each model could be analyzed in order to understand how differences in chemistries result in different sensitivities to the local environment embeddings defined by the spline filters.
The multi-element model trained to the datasets described in \Sec{flex_datasets} uses the same spline filters for all of the data, but separate NNs for each element type.
The network architecture maintained a funnel-like structure of $(12, 8, 4, 2, 1)$.
It also used four additional 3-body filters in order to increase its fitting capacity, for a total of one short-range pair filter and 12 3-body filters, though without the use of skip connections or the Softplus activation function.
In the case of the multi-element model, the use of skip connections would imply a ``background" energy that is consistent across all four element types and is augmented by the network contributions for each element.
While this assumption sounds plausible in theory, we found that in practice a model using skip connections resulted in poorer property predictions than one which did not.
This decreased performance when using skip connections can likely be attributed to the larger concentration of Al data dominating the fitting and causing the filters to learn a background energy that is influenced by the high-energy Al configurations and therefore not transferable to the lower energy Cu/Ge/Mo data.
We believe that further research into constructing balanced training datasets could help address this issue, though it may also be possible that a larger number of splines is required for describing this more diverse dataset.
Another alternative, which was not explored here, would be to only use skip connections for a subset of the spline filters in order to give the model the ability to learn a transferable background energy without enforcing the full constraint of skip connections for all filters.

The property predictions of the multi-element model plotted in \Fig{spider} show that a model using shared filters for multiple datasets can learn to make reasonable property predictions for all four elements studied in this work.
While the initial predictions for most of the elements were relatively good, the cubic elastic constants and bulk modulus for Mo were noticeably under-predicted, and the vacancy migration energy was far too large to be considered acceptable (dotted line in \Fig{spider}).
Due to the relatively few number of spline filters used, and their high degree of interpretablity, we were able to fine-tune the model by zeroing out the network weights of hand-chosen filters in order to remove their contribution to the Mo energy predictions.
For example, the contributions of each spline filter can be visualized individually in order to isolate the influence of each filter on the properties of interest (see \Fig{supp:mo_e_vs_a}).
Following this approach, we were able to improve the Mo predictions to bring them within an acceptable range without altering the predictions for the other three elements (dashed lines in \Fig{spider}).
However, we note that the original training errors of the model (before fine-tuning) as shown in \Fig{supp:multi_element_errors} were comparable to those of the NNIPs from \cite{Zuo2020}, suggesting that the property predictions of the multi-element s-NNP model may have been able to have been improved by re-balancing the dataset.
Similar to what was described in \Sec{flex_datasets}, a possible explanation for this is that the Mo dataset did not fully constrain the portions of the spline filters relevant to computing the properties of interest, and their shapes were therefore dictated by the Al dataset, leading to corrupted Mo property predictions.
Further research into methods for preventing this type of ``cross-pollution" of information would be valuable for constructing more flexible and generalizable models.

\begin{figure}[!ht]
    \centering
    \includegraphics[scale=0.625]{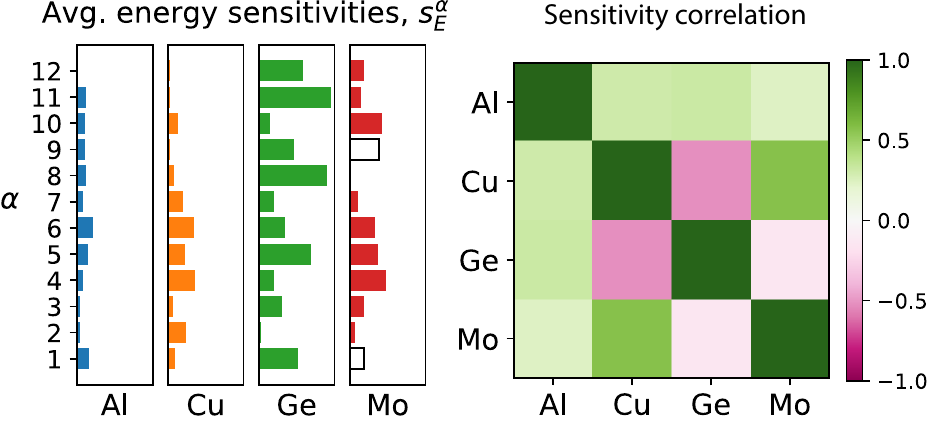}
    \caption{Analysis of sensitivities $s^\alpha_E$ (\Eqn{sensitivity}) of the predicted energies with respect to the 3-body spline filters, $G^\alpha_3$, for each elemental model. We compute the sensitivity by calculating the derivative of the predicted energy with respect to the spline filter outputs, normalized by the DFT reference value, then summing over all atoms in the training set.
    % The two models for systems with ground states of FCC (Al and Cu) or BCC (Mo) show similar sensitivities, with some variations in filters 3 and 7, while the diamond systems (Ge) shows significant differences across all filters.
    Note that these sensitivities are computed with respect to the total energy $E$, which includes the network contributions.
    The left panel plots the raw values, while the right panel shows the correlation between the sensitivities of each element.
    The empty bars for Mo correspond to the filters which were removed from the Mo energy predictions during fine-tuning as described in \Sec{flex}.
    The corresponding spline filters are visualized in \Fig{supp:filter_viz_multi_element}.
    }
    \label{fig:sensitivities}
\end{figure}

In order to gain additional insights into the differences in energetics between the Al, Cu, Ge, and Mo systems, we compute the sensitivities of the energies predicted by the multi-element model with respect to each of the 3-body spline filters. The sensitivity of the total energy, $E$, for all configurations $N$ with respect to a given 3-body spline filter $G^\alpha_{3}$ can be computed as:

\begin{equation}
    s^{\,\alpha}_E = \sum_{i=1}^{N}| \partial E_\text{s-NNP} / \partial G_{3,i}^\alpha | / E_\text{DFT},
    \label{eqn:sensitivity}
\end{equation}

\noindent
which can be easily computed via back-propagation.

Calculation and comparison of these sensitivities for each dataset, as shown in \Fig{sensitivities}, highlight the learned similarities and differences between the four elemental datasets.
Examination of the correlation between the sensitivities (right panel of \Fig{sensitivities}) shows that the Cu and Mo datasets have similar filter sensitivities, while Ge is the only element to have a negative correlation with any of the other elements.
The Al dataset appears to be uniformly similar to all three remaining elements, reflecting the fact that most of the geometric configurations of the Cu/Ge/Mo datasets are well-sampled by the Al dataset (see \Fig{supp:umaps}).
We hypothesize that the negative correlation of the Ge sensitivities is due to an attempt by the model to encode the large cluster of outlying Ge points in the UMAP visualizations shown in \Fig{supp:umaps}.

% These sensitivites can then be compared to the visualizations in \Fig{supp:filter_viz_multi_element} in order to understand the influence of specific physical features on the energy of the system.
% For example, \Fig{sensitivities} shows that all four models have a high sensitivity to filter number 5, which we see from \Fig{filter_viz} corresponds to a large repulsive contribution for small bond angles.

\section{Discussion} \label{sec:discussion}

\subsection{Understanding model behavior}
\label{sec:interp_discuss}

The main purpose behind the modifications proposed in \Fig{interp} and \Sec{interp} is to simplify the process of understanding the behavior of s-NNP models.
This not only improves the usefulness of visualizations like those shown in \Fig{filter_viz} and \Fig{supp:filter_viz_multi_element}, but can also aid in debugging model predictions in practice.
In this section, we will discuss each modification from \Sec{interp} in greater detail in order to highlight how they improve the interpretability of the model.

The first of these modifications is the use of skip connections, which result in strong theoretical changes that can not only lead to improved trainability as seen in \Tab{fits} and other deep learning applications \cite{Li2017}, but also greatly increase the interpretability of the model.
Modifying $\vec{G}_i$ to have units of energy seemingly contrasts with the notion of the atomic ``density", $n_i$, from \Eqn{smeam} and the local environment descriptor, $\vec{D}_i$, from \Eqn{nnp}, both of which are considered to be intermediate representations which will only become energies once they have been transformed by a regression function ($U_{c_i}$ or $N_{c_i}$ respectively).
However, we emphasize that these three quantities ($n_i$, $\vec{D}_i$, and $\vec{G}_i$) remain closely related even when $\vec{G}_i$ is in units of energy while the others are not.
Although local environment descriptors are usually seen as quantifying the geometry within a local neighborhood, there is no reason that the environment descriptor may not itself also be an energy.
An energy-based descriptor could then be thought of as a type of energy partitioning scheme, where in the case of s-NNP with skip connections the atomic energies contribute both linearly to the total energy (through the skip connection) and non-linearly (through the network).
In fact, in our previous work \cite{Vita2021} we observed that $U_{c_i}$ was often learned to be a nearly linear function, meaning that it essentially served the purpose of a simple unit conversion for $n_i$.
This linearity is essential to model interpretability, and is the main motivator for the use of skip connections in s-NNP, as it is a step towards simplifying the process of understanding how the spline filters contribute to the total energy.

Nevertheless, care must still be taken when interpreting the filter visualizations when the other modifications discussed in \Sec{interp} are not also employed.
For example, skip connections alone do \textit{not} guarantee that a negative filter value means $E_i$ will also be negative, since the network contribution $N(\vec{G}_i)$ may outweigh the skip connection term.
Similarly, adjusting the filter outputs (e.g., by tuning the knot positions) may have unexpected results due to the non-linear behavior of the network.
These issues are also present with all non energy-based descriptors like $n_i$ and $\vec{D}_i$, but can be further improved upon using the remaining techniques discussed in this section.

In order to further facilitate straightforward interpretation of spline visualizations, we also choose to pin the knots of the radial splines to be 0 at the cutoff distance $r_c$, and we remove bias terms from the layers of the NN.
As a result of this, both $N(\vec{G}_i)$ and $E_{\text{sc},i}$ will smoothly decay to 0 at the cutoff distance, thus incorporating the desired behavior which is enforced in the Behler-Parrinello NNP using the cutoff function $v_c$ in \Eqn{nnp_cutoff}.
In order to account for the fact that some datasets may have non-zero energy at the cutoff distance, an external bias term, $b$ is added so that the atomic energy converges to $b$ at $r_c$.
In this sense, $b$ corresponds to the learned energy of the isolated atom.
We note that this external bias term is preferable to using internal network bias or un-pinning the radial splines because it ensures that $N(\vec{G}_i)$ and $E_{\text{sc},i}$ behave similarly as $r_{ij}$ approaches $r_c$.
% Again, these modifications alone do not address the interpretability issues described in \Sec{skip}, but are rather an intermediate step that will help to isolate certain physical behavior to either the spline filters or the NN.

Wrapping the network output $N_\text{no-bias}(\vec{G}_i)$ in the Softplus activation function guarantees that the NN contributions to 
% \cite{Fukushima1975}
the total energy are strictly non-negative.
This ensures that any attractive behavior of the model arises solely from the spline filters through the skip connection term, $E_{\text{sc},i}$, thus helping to isolate certain model behaviors to specific parameter sets.
It is important to note, however, that the Softplus activation should only be used in conjunction with skip connections in order to ensure that the model can predict negative energies.

A common challenge for many MLIPs is ensuring a repulsive behavior at small values of $r_{ij}$.
This difficulty is not reflected in classical potentials, however, as classical models often include explicit repulsive terms.
While a data-driven solution to this problem is possible, by explicitly adding short-range dimers into the training dataset, many MLIP developers choose instead to adjust the form of their IP.
% As a result, it is relatively common for MLIP developers to either take the data-driven approach of adding short dimers to the training dataset.}
One such method used in the literature is to augment the model by introducing an auxiliary potential designed to capture the repulsive behavior of a pair potential.
For example, in \cite{Milardovich2023} a ``composite potential" was constructed by first fitting a repulsive ``auxiliary potential" to DFT data, then fitting a ``main potential" to the residuals of the pair potential using a Gaussian Approximation Potential \cite{Bartok2010}.
% Alternatively, the SNAP class of potentials \cite{Thompson2015} utilize a switching function which switches the IP to a repulsive analytical form at small distances \cite{Thompson2015}.
We incorporate a similar idea by including a short-range 2-body filter, $G_{2,i}$, which is summed directly into the total energy without ever passing through the NN.
Note that this is different from the skip connections, which are summed directly into the energy \textit{and} passed through the NN.
Having a spline filter which only contributes linearly to the total energy means that it is free of any of the interpretability issues described for the spline filters which are used as input to the NN.
While this short-range pair term does not necessarily guarantee a repulsive behavior at small distances, we observe empirically that it is often learned during training to have a strongly repulsive shape.
% Repulsive behavior could be more explicitly enforced by requiring the auxiliary filter to have a large negative slope at the innermost knot, however we found this to be unnecessary for the datasets used in this work.
% Interestingly, we observed that a short-ranged ($r_c < 3 \AA$) auxiliary 2-body filter could be trained in just a single epoch, and resulted in a sharp decrease in the loss function during the early stages of training.
A similar technique is used by the SNAP interatomic potential \cite{Wood2018}, where the repulsive ZBL pair potential is added to the potential form as a ``reference potential".

\subsection{s-NNP's relation to other models}
\label{sec:other_models}

The s-NNP architecture shown in \Fig{model} can be further understood by drawing relationships between itself and other models from the literature, particularly the UF3 model \cite{Xie2021} and the original Behler-Parrinello NNP \cite{Behler2007}.
%, and the MACE family of potentials \cite{Batatia2022}.
% The difference between s-NNP and these other models arises from either the embedding function used to encode the local atomic environment into a descriptor, or the regression function used to map the descriptor into an energy.
s-NNP can be compared to UF3 and NNP by analyzing the differences between the two key components of each model: the embedding function for encoding local atomic environments into a descriptor, and the regression function for mapping the descriptor into an energy.
Although it can be difficult to clearly distinguish between the embedding/regression portions of most MLIPs due to the inability to definitively establish the roles of all parameters in a deep model, we will attempt to break each model down in intuitive ways to facilitate comparison.
One can view the vector $\vec{G}_i$ from \Eqn{snnp} as a descriptor generated by an embedding function defined by the spline filters $G^\alpha_{3,i}$ and $G^\beta_{2,i}$.
This embedding technique is most similar to the UF3 method, which also decomposes the energy into two-body and three-body terms described by spline basis functions.
Though there are some differences in the exact details of the UF3 and s-NNP spline functions, for example UF3's use of tricubic B-splines instead of the 1D cubic Hermite splines of s-NNP, the general principle is the same.
While s-NNP's embedding function is most similar to that of UF3, its regression function is identical to that of the Behler-Parrinello NNP \cite{Behler2007}.
Therefore, in order to help analyze the performance of s-NNP with respect to other models in the literature, it is suitable to think of s-NNP as a combination of a UF3-like embedding function with an NNP regressor.
Or, equivalently, as an NNP using a spline-based embedding function instead of atom-centered symmetry functions \cite{Behler2007}.
However, neither UF3 nor NNP utilize all of the interpretability improvements discussed in \Sec{interp}.

The recently-proposed EAM-R model \cite{Nitol2023} is the most closely related model to s-NNP in the literature, as it also incorporates components of both a classical model (EAM) and an MLIP (an NNP).
EAM-R is a composite potential (using the terminology described in \Sec{interp_discuss}) where the auxiliary potential is an EAM model, and the main potential is an RANN (an NNP with descriptors inspired by the analytical MEAM equations \cite{Baskes1992}).
Similar to this work, the developers of EAM-R observed that combining a classical model with an MLIP resulted in both improved stability relative to an MLIP and improved accuracy to a classical potential alone.
Despite s-NNP's similarity to EAM-R, s-NNP has some key differences, namely the use of spline filters (instead of analytical MEAM-inspired descriptors) and some of the interpretability improvements discussed in \Sec{interp}.
In particular, the spline filters may be expected to be more flexible than the RANN descriptors (similar to how s-MEAM is more flexible than analytical MEAM) for a given computational cost, and benefit from the ability to enforce smoothness and convergence through curvature penalties and knot pinning.
Furthermore, s-NNP's use of skip connections and the removal of the internal network bias greatly improve the interpretablily of the model, as discussed in \Sec{interp}.
Although EAM-R does not include these modifications, it is an excellent example of an MLIP which could easily incorporate these same interpretability improvements.

\subsection{Computational costs}

The computational cost of s-NNP is dominated by the evaluation of $\vec{G}_i$, and is particularly dependant upon the choice of $N_3$.
In fact, basic profiling tests revealed that the filter evaluations accounted for approximately 95\% of the total CPU and GPU time.
This behavior can be understood heuristically by the fact that \Eqn{g} involves approximately $\mathcal{O}(N_n^2)$ spline evaluations for each filter where $N_n$ is the average number of neighbors within the cutoff distance (due to the summation over triplets of atoms), as opposed to the relatively few matrix multiplications associated with the evaluation of the neural network.
In general, the computational cost of inference with an s-NNP potential will scale sub-linearly with $N_3$ (some speedup can be achieved by performing batched spline evaluations for the filters in \Eqn{g}).
An important practical implication of this is that in order to improve the accuracy of a given s-NNP model (and other NNP-based MLIPs as well), it is much more computationally efficient to increase the size of the network rather than the size of the embedding function.
On the other hand, increasing $N_3$ may lead to larger increases in accuracy (up to a point) than what is achievable by only increasing the network size

Given the performances of the s-NNP models in \Tab{fits} and the timing comparisons observed in our previous work between s-MEAM and a NNP \cite{Vita2021}, it may be expected that an s-NNP could be constructed that achieves identical errors to ANI while maintaining a higher speed.
The ``(1,8) $l=5$, int. all" model is already nearing this threshold, especially taking into account that the values for ANI reported in \Tab{fits} use ensemble-averages over 8 models, as reported by the original authors \cite{Smith2021_al_al}, which they say can make the energy and force errors ``20\% and 40\% smaller, respectively" and may account for the differences in performance as compared to ``(1,8) $l=5$, int. all".

\section{Conclusion}

In this work we developed a novel framework using spline-based filters coupled with a neural network regressor in order to blend the strengths of both classical and ML IPs.
We use this framework to probe the gap between these two classes of models, observing performance limits of linear (``classical") models that can be overcome using even small neural networks.
We then show that this improved performance can be maintained while incorporating architectural changes which improve the interpretability of the model, such as the use of skip connections, an external bias term, a Softplus activation function, and a short-range pair term.
Finally, we demonstrate that the information-rich filter layer can be used as a reference point for performing cross-system analyses, and correlate well enough with model behavior to enable manual tuning to improve property predictions.
Future studies applying the visualization techniques shown here to practical applications, and exploring methods of isolating contributions of specific elements to subsets of the model parameters would be valuable for continuing to refine the interpretability improvements explored in this work.
Furthermore, efforts building upon this work could continue to improve the s-NNP design by incorporating equivariance or a message-passing network, which may lead to better performance and improved scaling of model size with number of elements.

% \clearpage

\section{Code and Data Availability}
The code used for training s-NNP models can be found at \url{https://github.com/TrinkleGroup/snnp}. All datasets can be obtained from their original sources.

\section{Acknowledgements}
This research was supported by the National Science Foundation through awards NSF/NRT-1922758 and NSF/HDR-1940303. Computational resources were provided on the HAL computing cluster \cite{Kindratenko2020} by the National Center for Supercomputing Applications.

\bibliography{bibliography}

% \clearpage
\appendix
% \onecolumn

% Supplementary information

\section{Training details}
\label{sec:appendix:training}

All s-NNP models in this work were trained using the code provided at \url{https://github.com/TrinkleGroup/snnp} on a single GPU from the HAL computing cluster \cite{Kindratenko2020}. The energy and force terms in the loss function were given weights of 10 and 1 respectively. L2 regularization was applied to the network parameters with a weight of 0.01. The AMSGrad variant of the Adam optimizer was used, with an initial learning rate of 0.001 and a MultiStepLR scheduler. Inner and outer cutoff radii of 2.5 and 7.0 were used, as specified in the main text.

\section{Distributions of $r_{ij}$ and $\theta_{jik}$}
\label{sec:appendix:rij_theta}
\renewcommand{\thefigure}{B\arabic{figure}}
\setcounter{figure}{0}
\renewcommand{\thetable}{B\arabic{table}}
\setcounter{table}{0}

\begin{figure}[!ht]
    \centering
    \includegraphics[scale=0.4]{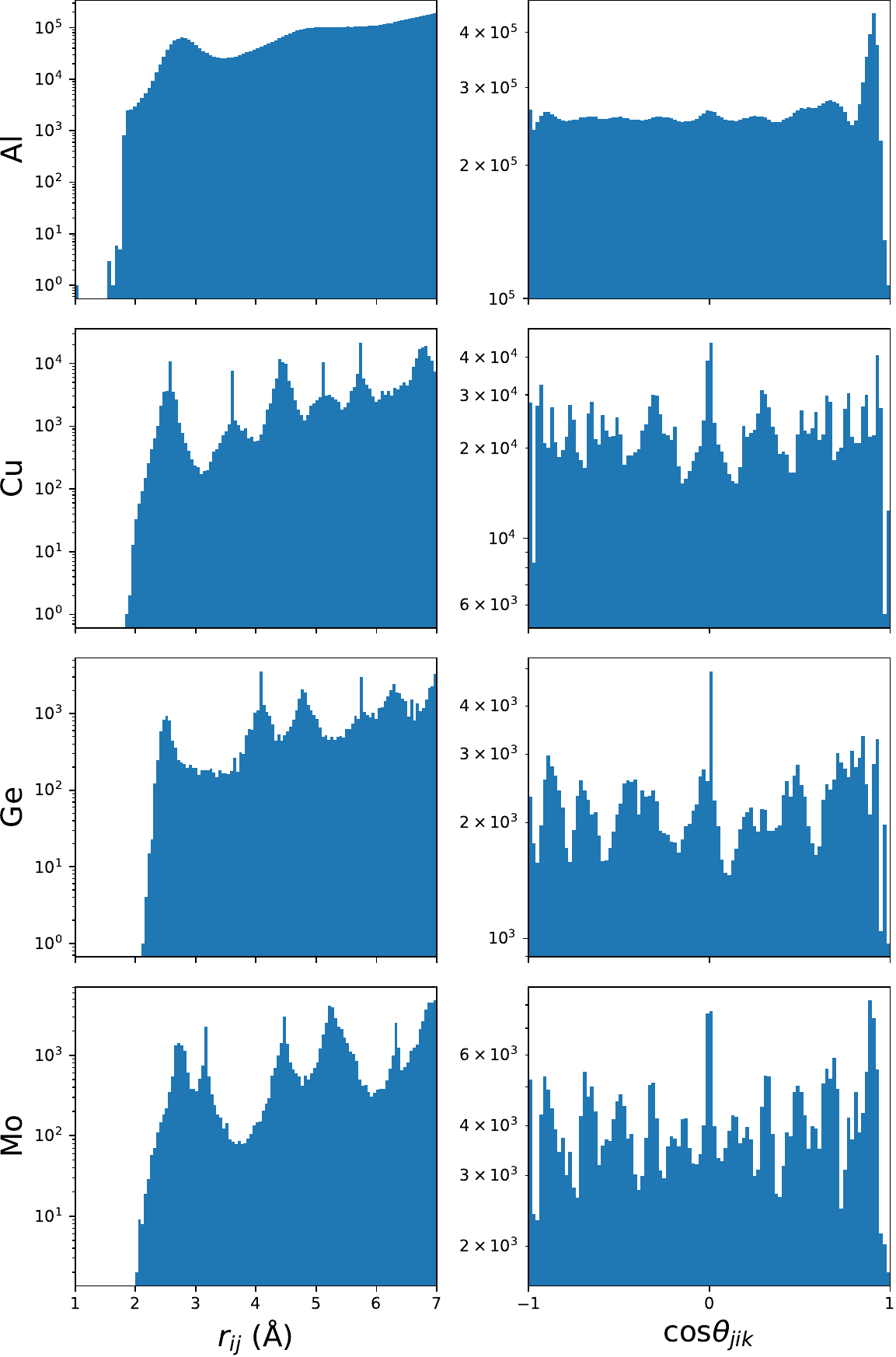}
    \caption{Histograms of $r_{ij}$ and $\cos{\theta_{jik}}$ values sampled from the Al, Cu, Ge, and Mo datasets used in this work.}
    \label{fig:supp:rij_costheta_distributions}
\end{figure}

\Fig{supp:rij_costheta_distributions} shows that the Al dataset well samples the range of $r_{ij}$ and $\cos{\theta_{jik}}$ values present in the Cu, Ge, and Mo datasets. Notably, the Al dataset has a much more uniform sampling of both of these values, likely due to both its large size and more diverse sampling technique (active learning). These results suggest that the Al dataset may help to constrain regions of the 2-body and 3-body splines that are under-sampled by the other three datasets. The inner cutoff radius of 2.5 $\AA$ (below which the short-range pair term becomes non-zero) was chosen to be slightly larger than the smallest atomic distances sampled by the datasets in order to ensure that the inner knots of the splines would be well constrained by the data.

\section{Multi-element model}
\renewcommand{\thefigure}{C\arabic{figure}}
\setcounter{figure}{0}
\renewcommand{\thetable}{C\arabic{table}}
\setcounter{table}{0}

\begin{figure}[!ht]
    \centering
    \includegraphics[width=\linewidth]{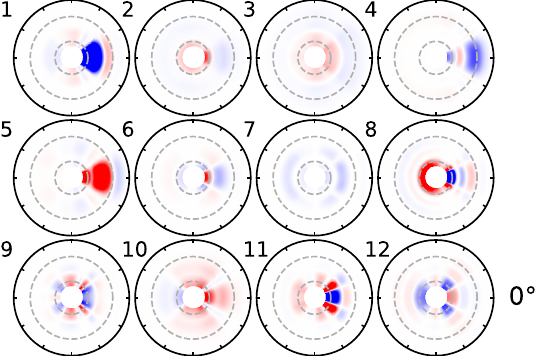}
    \caption{Polar plot visualizations of the 3-body spline filters for the multi-element s-NNP. Grey dashed lines mark 2\AA \space and 5\AA \space for reference.}
    \label{fig:supp:filter_viz_multi_element}
\end{figure}

\begin{figure}[!ht]
    \centering
    \includegraphics[width=\linewidth]{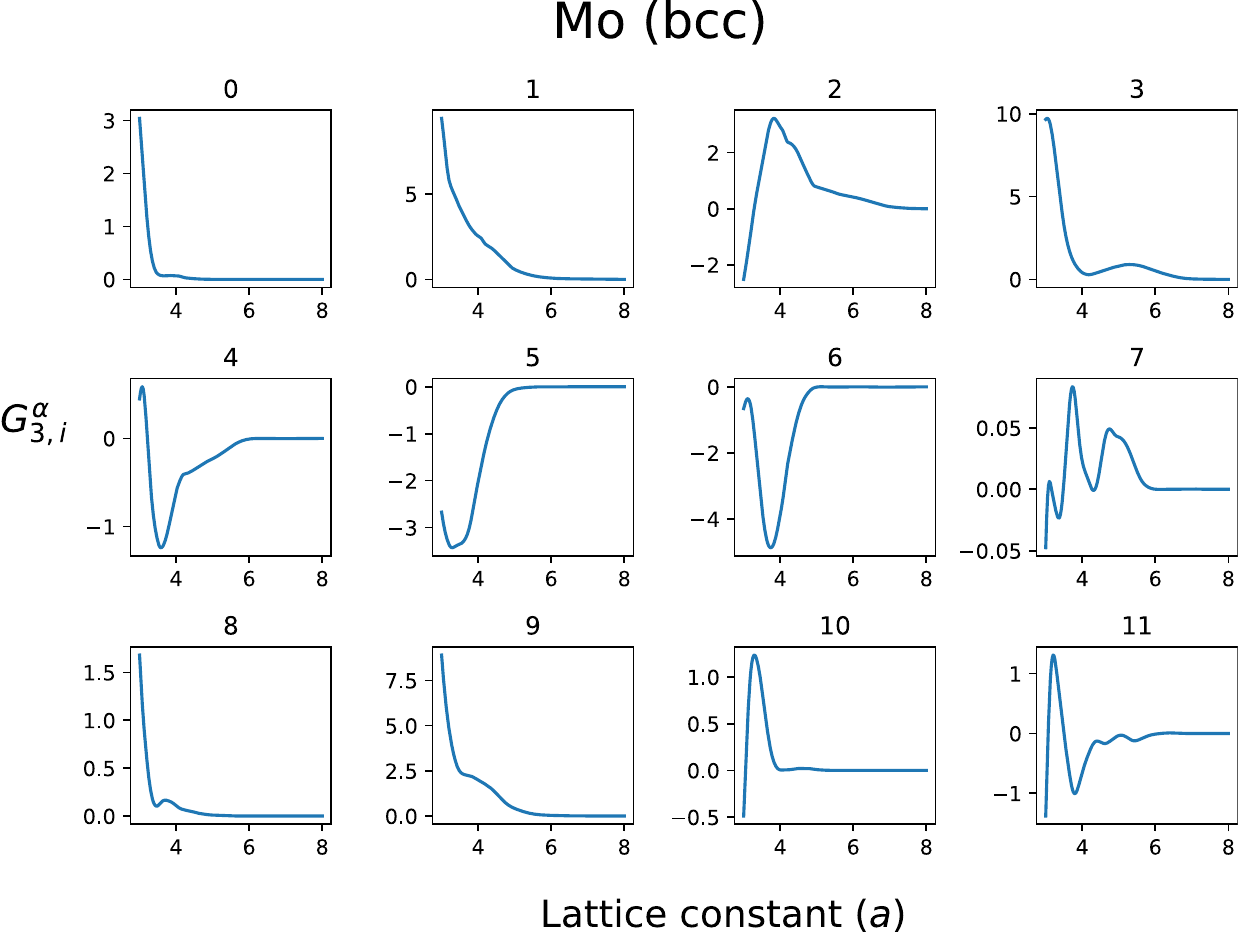}
    \caption{Spline filter activations, $G_{3,i}^\alpha$, for the multi-element model from \Sec{flex} on the $E$ vs. $a$ curve for BCC Mo.
    Filter numbers 1 and 9 were removed from the Mo property predictions during the fine-tuning process by zeroing out the corresponding network weights as described in \Sec{flex}.
    }
    \label{fig:supp:mo_e_vs_a}
\end{figure}

\begin{figure}[!ht]
    \centering
    \includegraphics[width=\linewidth
]{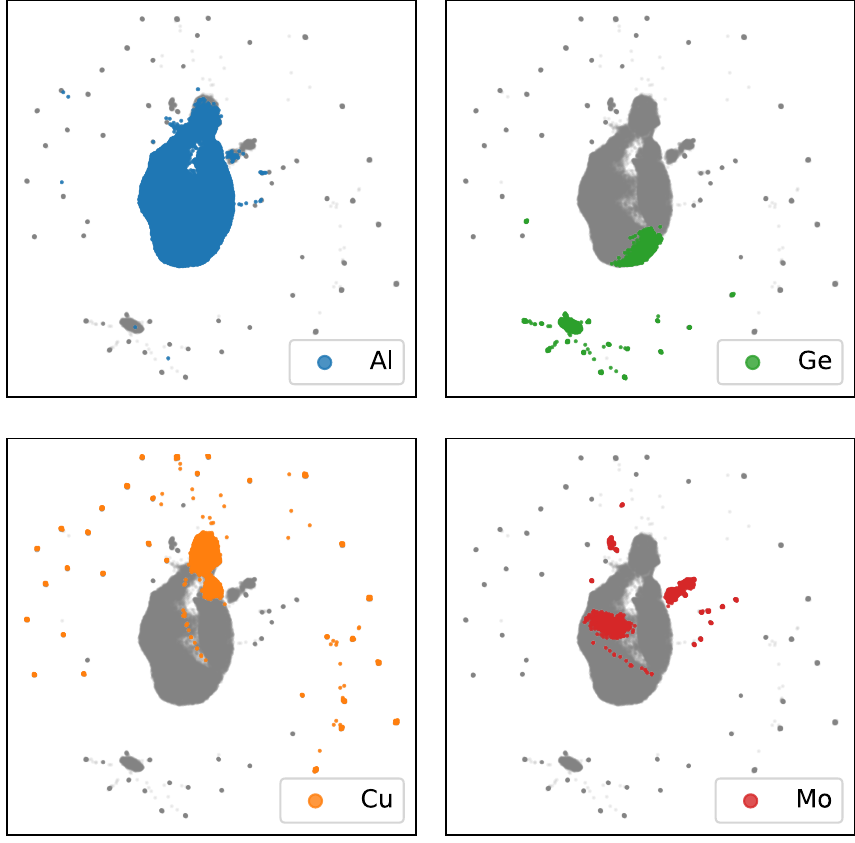}
    \caption{UMAP plots of the  vectors $\vec{G_i}$ generated by applying the spline filters from the multi-element s-NNP model from \Sec{flex}, colored by element type.
    % Since the $\vec{G_i}$ are per-atom quantities, whereas the DFT energies are supercell energies, each $\vec{G_i}$ from a given supercell is colored according to the per-atom energy of that supercell. While this energy partitioning likely does not reflect the true partitioning of the system, it provides a simple method of approximating the true energy distribution of the latent space containing the $\vec{G_i}$. An alternative method would be to sum $\vec{G_i}$ over all atoms in a supercell, however this would obscure the fact that some of the local atomic environments from the Cu/Ge/Mo datasets are well-represented in the Al dataset even if the supercells are not.
    }
    \label{fig:supp:umaps}
\end{figure}

\begin{figure}[!ht]
    \centering
    \includegraphics[scale=0.5]{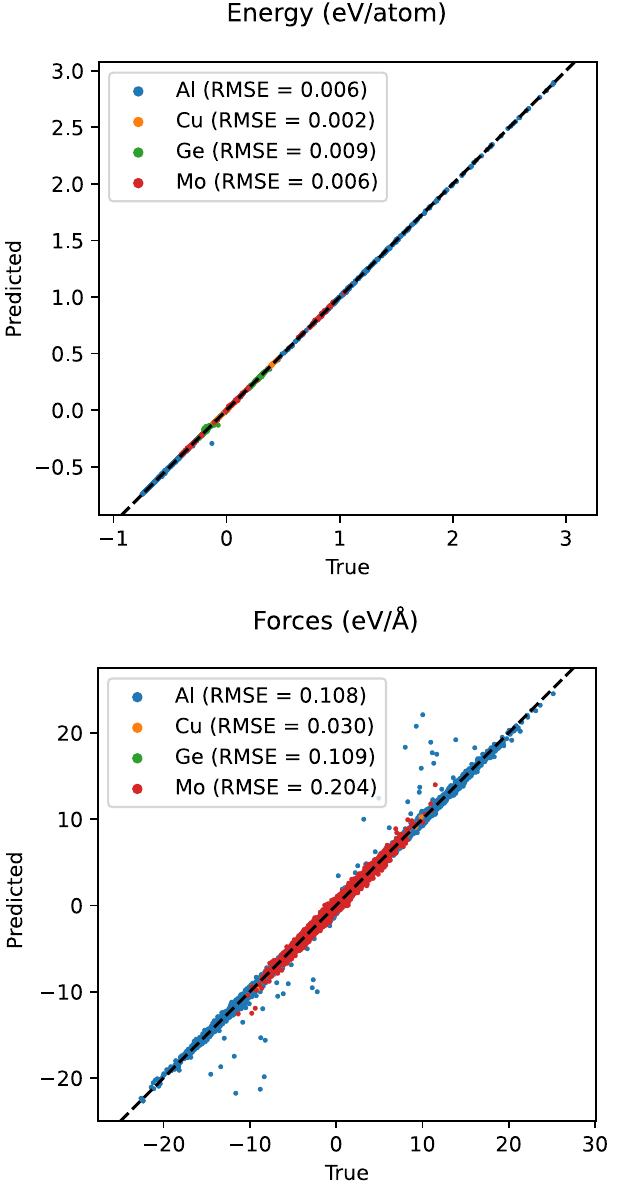}
    \caption{Parity plots for energy and force predictions of the multi-element model described in \Sec{flex} before manual fine-tuning of the model.
    The computed RMSE values match the expected errors for NN-based MLIPs on the Cu/Ge/Mo dataset based on the results from \cite{Zuo2020}, though the s-NNP model here uses a larger cutoff distance.
    Note that the RMSE value for Al should not be compared directly to those in \Tab{fits}, since the results here use only the 20\% of the Al dataset as described in \Sec{flex_datasets}.
    }
    \label{fig:supp:multi_element_errors}
\end{figure}

\Fig{supp:filter_viz_multi_element} shows the spline filter visualizations corresponding to the 12 splines used by the multi-element model. As described in \Sec{flex}, the multi-element model does not use skip connections, which means that the filters in \Fig{supp:filter_viz_multi_element} are not necessarily in units of energy and may be drastically transformed before being mapped into the output of the model. Although this greatly reduces the utility of the spline visualizations, the filters can still be understood in the context of the model sensitivities reported in \Fig{sensitivities}. It can be seen that filters 1 and 9 (which were removed during the manual fine-tuning process) both have negative activations for short bond lengths and bond angles less than $30 \degree$. The filters can be further analyzed by plotting their activations over a range of lattice constants, as shown in \Fig{supp:mo_e_vs_a}, which reveals that filters 1 and 9 both yield strongly repulsive contributions for small lattice constants. The removal of these filters for the Mo predictions drastically lowered the predicted $E_m$, which is consistent with our intuition that removal of filters 1 and 9 should result in a softer potential.

\Fig{supp:umaps} supports the expectation from \Sec{appendix:rij_theta} that the Al dataset encompasses a majority of the data from the Cu/Ge/Mo datasets. As can be seen from \Fig{supp:umaps}, this statement appears to be generally true, with the exceptions of some small outlying clusters of Cu and Ge points. The small isolated clusters surrounding the main manifold correspond to the strained configurations from the Cu/Ge/Mo datasets, while the larger outlying cluster of Ge points correspond to a subset of the Ge configurations which were sampled by low- and high-temperature MD simulations.

\Fig{supp:multi_element_errors} provides a breakdown of the errors of the multi-element model (prior to fine-tuning) for each elemental datasets. The apparent contradiction between the low RMSE values (compared to those from \cite{Zuo2020}) shown in \Fig{supp:multi_element_errors} and the unacceptably high errors in property predictions for Mo reported in \Fig{spider} suggests that the configurations in the Mo dataset were not sufficiently representative of the properties of interest. While it is possible that the property predictions of the multi-element s-NNP may have been able to be improved if even lower Mo RMSE values could have been obtained, the fact that many models with similar errors (\cite{Zuo2020} and \cite{Vita2021}) had good property predictions suggests some kind of deficiency in the dataset which would require further analysis in order understand fully.

\end{document}